\documentclass[,final,numberedheadings]{aipproc}

\layoutstyle{8x11single}

\begin{document}

\title{Pairing schemes for HFB calculations of nuclei}

\author{T. Duguet}{
  address={NSCL and Department of Physics and Astronomy, Michigan State University, East-Lansing, MI 48824, USA} }

\author{K. Bennaceur}{
  address={IPNL, CNRS-IN2P3 / Universit\'e Claude Bernard Lyon
1, 69622 Villeurbanne Cedex, France} }

\author{P. Bonche}{
  address={SPhT, CEA Saclay, 91191 Gif sur Yvette Cedex, France}
}

\begin{abstract}
Several pairing schemes currently used to describe superfluid
nuclei through Hartree-Fock-Bogolyubov (HFB) calculations are
briefly reviewed. We put a particular emphasis on the
regularization recipes used in connection with zero-range forces
and on the density dependence which usually complement their
definition. Regarding the chosen regularization process, the goal
is not only to identify the impact it may or may not have on
pairing properties of nuclei through spherical 1D HFB calculations
but also to assess its tractability for systematic axial 2D and 3D
mean-field and beyond-mean-field calculations.
\end{abstract}

\maketitle

\section{Introduction}
\label{intro}

It has been known for a long time that the structure of the
nucleus depends significantly on their superfluid nature. Indeed,
pairing constitutes the main part of the residual interaction
beyond the Hartree-Fock (HF) approximation and has a strong
influence on most low-energy properties of the
system~\cite{bender03b}. This encompasses masses, separation
energies, deformation, individual excitation spectra and
collective excitation modes such as rotation or vibration. The
role of pairing correlations is particularly emphasized when going
toward the neutron drip-line because of the proximity of the Fermi
surface to the single-particle continuum. Indeed, the scattering
of virtual pairs into the continuum gives rise to a variety of new
phenomena in ground and excited states of nuclei~\cite{doba3}.

Despite its major role, our knowledge of the pairing force and of
the nature of pairing correlations in nuclei is rather poor. The
importance of resolving the range of the interaction, the nature
and characteristics of its density dependence (in particular the
isovector and low-density parts) have to be
clarified~\cite{doba3,duguet2,doba4,bulgac3}. Also, the impact of
the Coulomb force~\cite{anguiano01a} and of the three-body
force~\cite{fayans00a,zuo02c} on pairing properties of nuclei is
not well understood. Last but not least, particularly puzzling is
the situation regarding beyond-mean-field effects. Screening
effects due to density and spin fluctuations are known to strongly
decrease the pairing gap in infinite neutron matter, both for
singlet and triplet
pairing~\cite{dean03a,clark76a,baldo02a,shen03a,schulze96a}.
Whether one can extend this result to finite nuclei is still an
open question. In particular, the induced interaction and
off-shell self-energy effects due to the exchange of surface
vibrations between time-reversed states seems to increase the
pairing gap in finite nuclei compared to that generated by the
bare force only~\cite{terasaki1,barranco04a}. In addition, the
influence of the restoration of particle-number and pairing
vibrations in even and odd nuclei still has to be characterized
through systematic calculations.

In the present paper, we discuss results of spherical 1D
Hartree-Fock-Bogolyubov (HFB) calculations using currently popular
pairing schemes. This set of results constitutes a preview of a
more extensive study~\cite{duguet06a}. Our goal at this stage is
not to discuss their relevance in terms of reproduction of
experimental data but rather to clarify their theoretical content
to make such an ultimate comparison meaningful. Also, their
tractibility for systematic 2D and 3D mean-field and
beyond-mean-field calculations is adressed. In
section~\ref{formalism}, we briefly outline the characteristics of
the different pairing schemes. Results are displayed in
section~\ref{results}. Conclusions are given in
section~\ref{conclusions}.

\section{Existing pairing schemes}
\label{formalism}

To treat pairing, one needs to specify the many-body technique
used and the appropriate interaction to insert into the
calculation at the chosen level of approximation. The latter
depends both on the situation and the system. In the present case,
we concentrate on a (self-consistent) mean-field description of
finite nuclei using the HFB method~\cite{bender03b}. Within such a
framework, we compare pairing schemes differing not only by their
analytical structure but also from the point of view of the
motivation for their adjustment/definition. Phenomenological
forces adjusted in the context of Density Functional Theory (DFT)
and supposedly renormalizing all possible correlations through HFB
calculations are compared to a microscopic vertex equivalent to
the bare nucleon-nucleon force. The latter scheme is motivated by
perturbative methods showing unambiguously that the interaction to
be used in the particle-particle (p-p) channel at lowest order in
irreducible vertices ({\em i.e.} well-defined mean-field theory)
is the bare nucleon-nucleon ($NN$)
force~\cite{gorkov,bogo,mehta1,henley}. At the next order, the
irreducible pairing vertex involves the so-called polarization
diagrams.

Until recently, only phenomenological pairing forces such as the
Gogny force~\cite{decharge80a} or (Density-Dependent) Delta
Interactions ((DD)DIs)~\cite{duguet2,doba4,bertsch,rigol} have
been used in mean-field and configuration mixing
calculations~\cite{bender03b}. Although successful in describing
low-energy nuclear structure over the {\it known} mass table, they
lack a clear link to the bare interaction. This feature strongly
limits the reliability of their analytical structure such as their
possible density dependence. Also, their direct fit to nuclear
data through mean-field calculations makes probable the
re-normalization of beyond-mean-field effects. This is a
significant limitation if, and only if, one wants to go explicitly
beyond that level of approximation. Finally, their fits performed
onto very limited sets of nuclei around stability make their
extrapolated use toward the drip-lines unsafe.

Let us now concentrate on DDDIs. The corresponding ansatz read as:

\begin{equation}
D \left(\vec{r}_{1} \sigma_{1}, \, \vec{r}_{2} \sigma_{2}\right)
\, = \, \lambda \, \, \frac{1 - P_{\sigma}}{2} \, \, f
\left(\vec{r}_{1}\right) \, \, \delta (\vec{r}_{1}-\vec{r}_{2}) \,
\, \, \, \, , \label{DDDI}
\end{equation}
where $P_{\sigma}$ is the spin-exchange operator,
$\rho_{0}(\vec{r})$ is the local scalar-isoscalar part of the
density matrix and $f \left(\vec{r}_{1}\right) = \left[1 -
\rho_{0}(\vec{r}_{1}) / \rho_{c} \, \right]$ is the
density-dependent form-factor. In the present case, we consider,
"surface-type" ($\rho_{c}=\rho_{sat}$, where $\rho_{sat}$ is the
saturation density of symmetric nuclear matter) and
"half-volume/half-surface type" ($\rho_{c}= 2\rho_{sat}$) DDDI. It
is worth noting that recent analysis of asymptotic matter and pair
densities of exotic nuclei~\cite{doba4}, the evolution of the
pairing gap toward the neutron drip-line~\cite{doba2}, moment of
inertia of transfermium isotopes~\cite{duguet2} and the average
behavior of the odd-even mass differences over the mass
table~\cite{doba7} have recently shown that an optimal
compatibility between experimental data and results of HFB
calculations was obtained for a DDDI between surface and volume.

When using DDDI to describe pairing, one needs to regularize the
theory which otherwise diverges~\cite{bulgac1}. There are several
phenomenological ways to do so. One scheme used in connection with
the two-basis method to solve HFB equations consists of limiting
the sum in the calculation of the pairing-field matrix elements to
single-particle states whose energies lie inside a (smoothed)
window $\lambda - 5 \leq \epsilon_{m} \leq \lambda + 5 $ around
the Fermi energy~\cite{gall}. Another popular scheme used when
solving the HFB problem in coordinate space consists of
calculating the densities by limiting the integrals performed in
the quasi-particle basis to states with $E_{qp} < E_{c} = 60$
MeV~\cite{doba84a,bennaceur05a}~\footnote{It is in fact in the
so-called "single-particle equivalent spectrum" which allows the
separation between particles and holes~\cite{bennaceur05a}.}.

It is only recently that the regularization of the pairing problem
could be understood microscopically~\cite{bulgac1,bulgac2}. The
idea is to identify the divergences stemming from the use of a
local gap and to regularize them through a well-defined
renormalization procedure. This complements the DDDI with an
effective coupling constant $g_\mathrm{eff} (\vec{r})$ expressed
in terms of an energy cut-off which has to be taken sufficiently
far away from the Fermi energy to ensure that observable are
independent of its value~\footnote{This parameter will be taken as
the canonical basis size $E_{max}$ in section~\ref{results} and in
Fig.~\ref{convener} when we test the convergence properties of the
different methods.}. Note that $g_\mathrm{eff} (\vec{r})$ comes in
addition to the genuine density-dependent form factor $f
\left(\vec{r}\right)$.

In an attempt to describe pairing in the $^{1}S_0$ channel
starting from the bare $NN$ interaction, a tractable form of the
latter for HFB calculations of finite nuclei was proposed
recently~\cite{duguet04a}. BCS pairing properties provided in
infinite matter by $AV18$~\cite{wiringa} were reproduced very
accurately. An effective version of the interaction (FR) was then
introduced through a recast of the gap equation into a fully
equivalent pairing problem. Resumming high energy pair scattering
into the effective vertex provided the latter with a factor $2 \,
v^{2}_{m}$, where $v^{2}_{m}$ denotes usual BCS occupation
numbers. Such a factor acts as a microscopic regularizator and
makes zero-range approximations meaningful. Thus, a zero-range
(ZFR) approximation providing in infinite matter the same gaps at
the Fermi surface as the bare force was defined~\cite{duguet04a}.

Beyond the regularization factor, the recast of the pairing
problem provides the effective pairing forces FR and ZFR with a
density dependence of the form~\cite{duguet04a}:

\begin{equation}
f_{(Z)FR} (\vec{r})  \, = \, A_{(Z)FR} + B_{(Z)FR} \, \ln
\rho_{q}(\vec{r} \,) + C_{(Z)FR} \, \left[\ln
\rho_{q}(\vec{r})\right]^{2} \, \, \, \, , \label{effectiveforce1}
\end{equation}
where $\rho_{q}(\vec{r})$ is the local density of nucleons of
isospin $q$. The isovector character of such density dependence is
different from the one introduced for usual phenomenological DDDI.
The coefficients entering the functional $f_{ZFR}
\left(\vec{r}\right)$ differ from those used for FR. The
surface-enhanced character of phenomenologically optimized
DDDI~\cite{duguet2,doba4,doba2,doba7} was {\it derived} in
Ref.~\cite{duguet04a} and shown to be, to a large extent, a way of
re-normalizing the range of the interaction. It was also shown
that usual DDDIs miss the low-density behavior of the effective
pairing force. This feature was briefly discussed along with the
first published results of 3D HFB calculations with FR and
ZFR~\cite{duguet05b}.

For the present study, we select/define a set of 6 representative
pairing schemes whose characteristics are summarized in
table~\ref{forces}. They combined the different possibilities
discussed previously regarding the range, density dependences and
regularization schemes. ULB is a commonly used surface-peaked
interaction adjusted on superdeformed bands of medium mass
nuclei~\cite{rigol}. DFTS, DFTM and RDFTM were adjusted in a
similar fashion as was done in Ref.~\cite{doba04a} by asking for
the average neutron gap in $^{120}Sn$ to be $< \Delta^{n}
(^{120}Sn) >_{\kappa} = \sum_{m>0} \, \Delta^{n}_{m} \,
\kappa^{n}_{m\bar{m}} \, / \, \sum_{m>0} \kappa^{n}_{m\bar{m}} =
1.245$~MeV. The definition of (Z)FR, free from any adjustment in
finite nuclei, has already been discussed. The Gogny force is
listed for completeness only.

\begin{table}
\label{forces}
\begin{tabular}{|c|c|l|c|c|}
\hline Name & Range & Regularization scheme & Density
form factor & Type of density \\
\hline \hline ULB & zero & phenom. $\lambda - 5 \leq
\epsilon_{m} \leq \lambda + 5  $ & $\rho_{c} = \rho_{sat}$ & $\rho_{0} (\vec{r})$ \\
\hline DFTS  &  zero & phenom. $E_{qp} \leq 60$~MeV &   $\rho_{c} = \rho_{sat}$    & $\rho_{0} (\vec{r})$ \\
\hline DFTM  &  zero & phenom. $E_{qp} \leq 60$~MeV &   $\rho_{c} = 2 \rho_{sat}$    & $\rho_{0} (\vec{r})$ \\
\hline RDFTM  &  zero & micro. $g_\mathrm{eff} (\vec{r})$
&    $\rho_{c} = 2 \rho_{sat}$    & $\rho_{0} (\vec{r})$ \\
\hline ZFR$^{\S}$  &  zero &  micro. $2 v^{2}_{i}$ & Surface
enhanced +
rising at low dens  & $\rho_{q} (\vec{r})$ \\
\hline FR$^{\S \dag \ast}$  & finite &
& &  \\
\hline
Gogny$^{\S}$  &  finite &  &   & \\
\hline
\end{tabular}
\caption{Pairing schemes currently used in HFB calculations of
finite nuclei. $^{\S}$ Incorporate the low-density behavior
associated with the virtual n-n state in the vacuum. $^{\dag}$ Is
explicitly connected to the bare $NN$ force in the $^{1}S_{0}$
channel. $^{\ast}$ FR is an effective form of the exact bare force
obtained through the recast of the gap equation. As a result, FR
incorporates the cut-off $2 v^{2}_{m}$ and the density form-factor
as defined by Eq.~\ref{effectiveforce1}. Of course, FR's finite
range makes the factor $2 v^{2}_{m}$ unnecessary as far as
convergence is concerned.}
\end{table}

\section{Results}
\label{results}

We perform 1D spherical HFB calculations of tin even-even
isotopes' ground-states. The HFB equations are solved using the
two-basis method~\cite{gall}. The Sly5 Skyrme
force~\cite{chabanat} is used in the p-h channel. The calculations
are re-done for each pairing scheme listed in table~\ref{forces},
except for the Gogny force. By keeping the same force in the p-h
channel, the pairing vertices are probed and compared in a
consistent manner, including the self-consistent coupling between
the two channels. Of course, properties of the force in the p-h
channel have an impact on the results. In that respect, it is
worth noting that the considered DDDI were adjusted together with
the SLy5 (SLy4 for ULB) parameterization in the p-h channel whose
isoscalar effective mass is $m^{\ast}/m = 0.7 $. Also, $ZFR$ and
$FR$ were defined once for all without any reference to finite
nuclei and are insensitive to the effective mass used as far as
their parameters values are concerned~\cite{duguet04a}.

\begin{figure}
  \includegraphics[angle=0,scale=0.42]{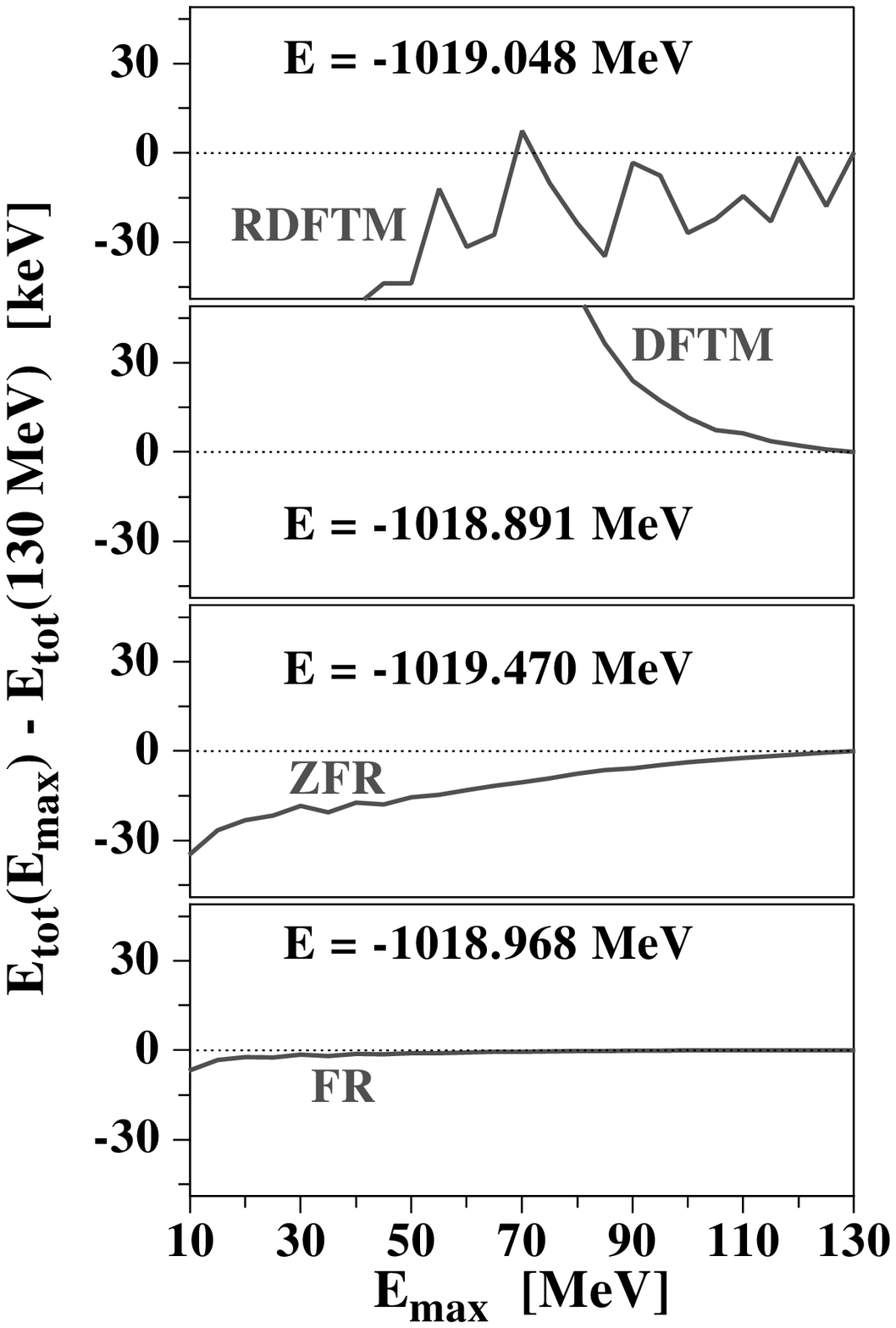} \hspace{-2.5cm}
  \includegraphics[angle=0,scale=0.42]{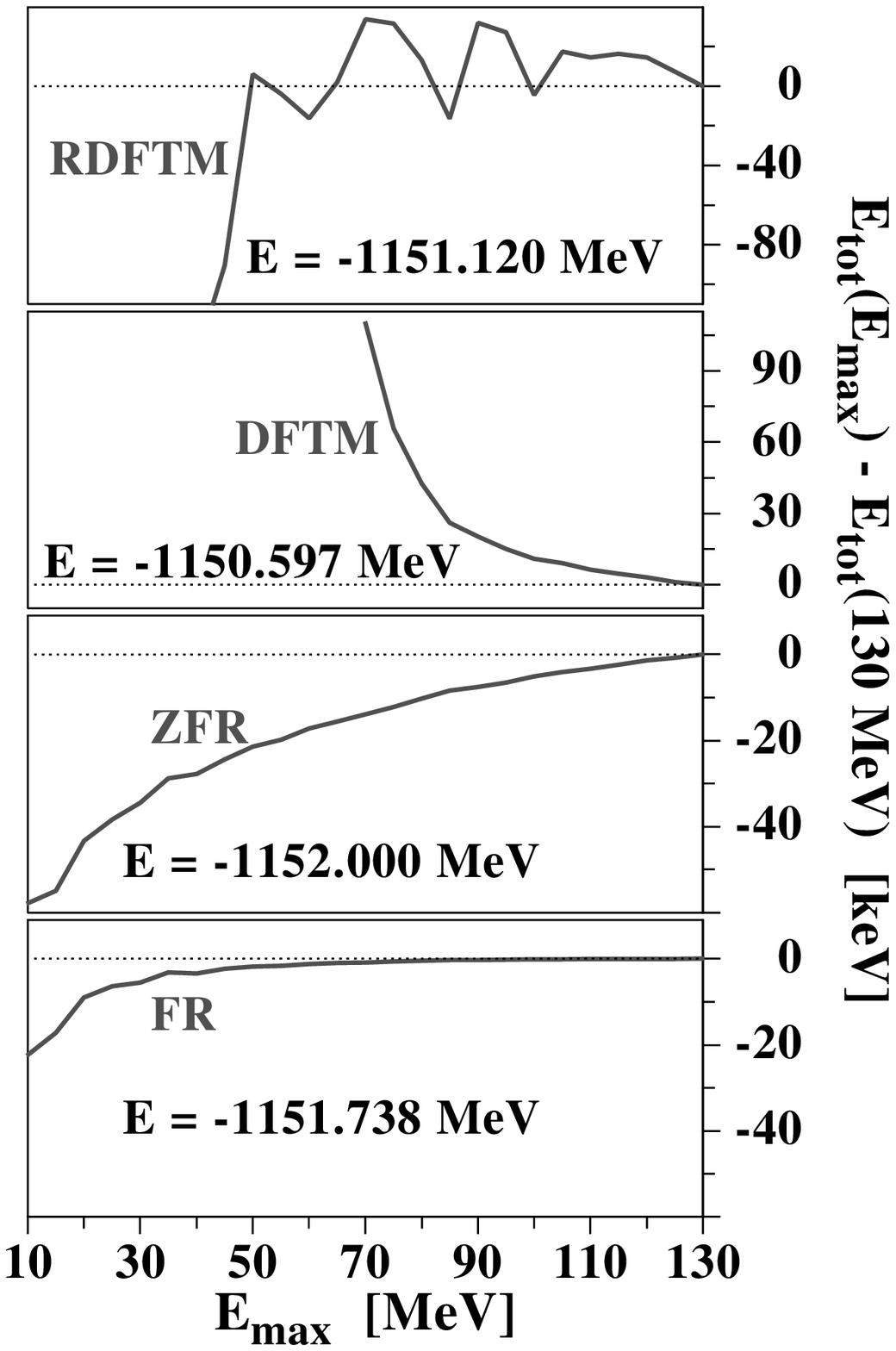}
  \caption{Left panel: binding energy of $^{120}Sn$
   as a function of the energy
$E_{max}$ of the highest single-particle state included in the
canonical basis. Right Panel: same as left panel for $^{170}Sn$
(vertical scales are different).}
  \label{convener}
\end{figure}

\begin{figure}
  \includegraphics[angle=270,scale=0.32]{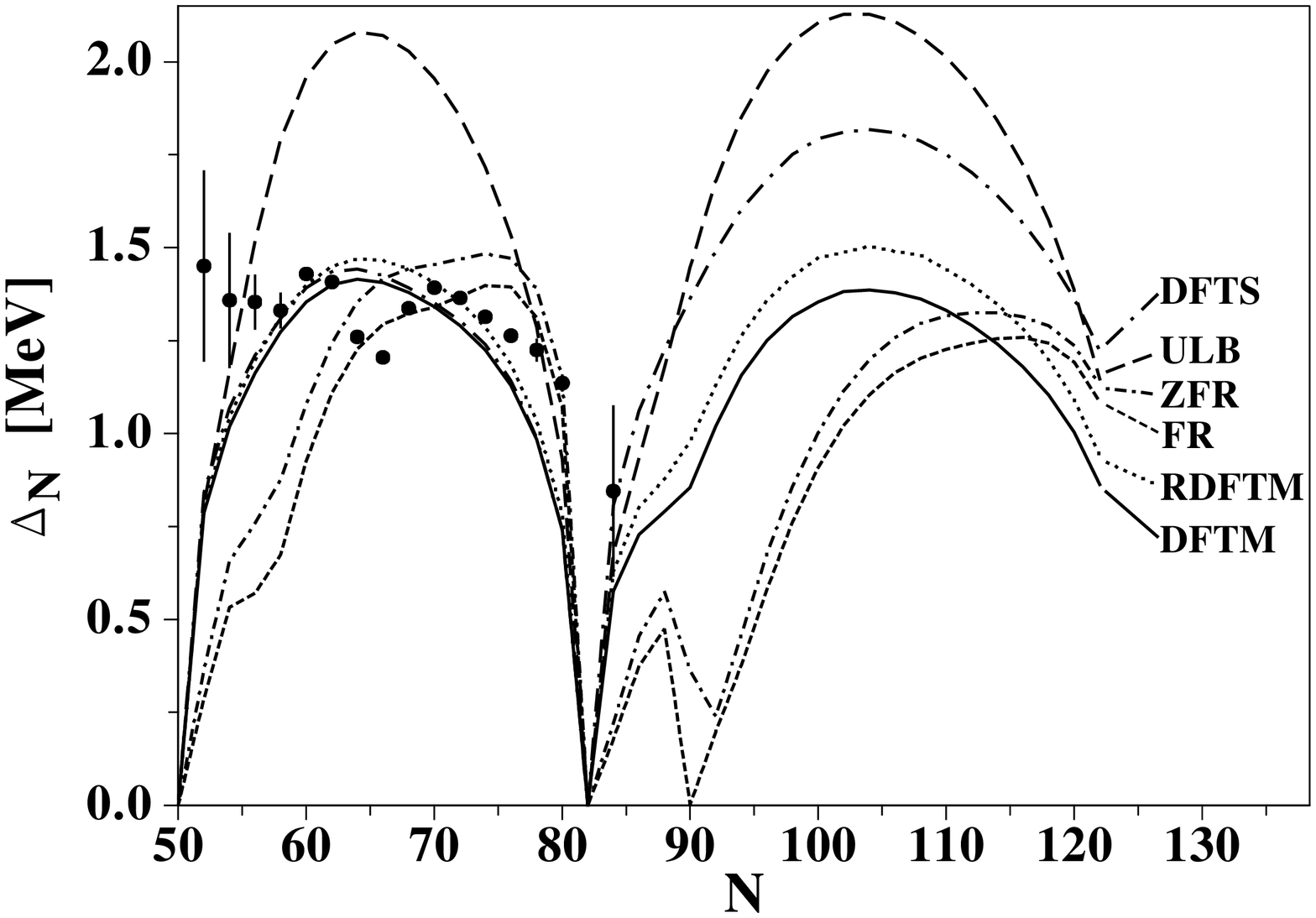} \hspace{-2cm}
  \includegraphics[angle=270,scale=0.32]{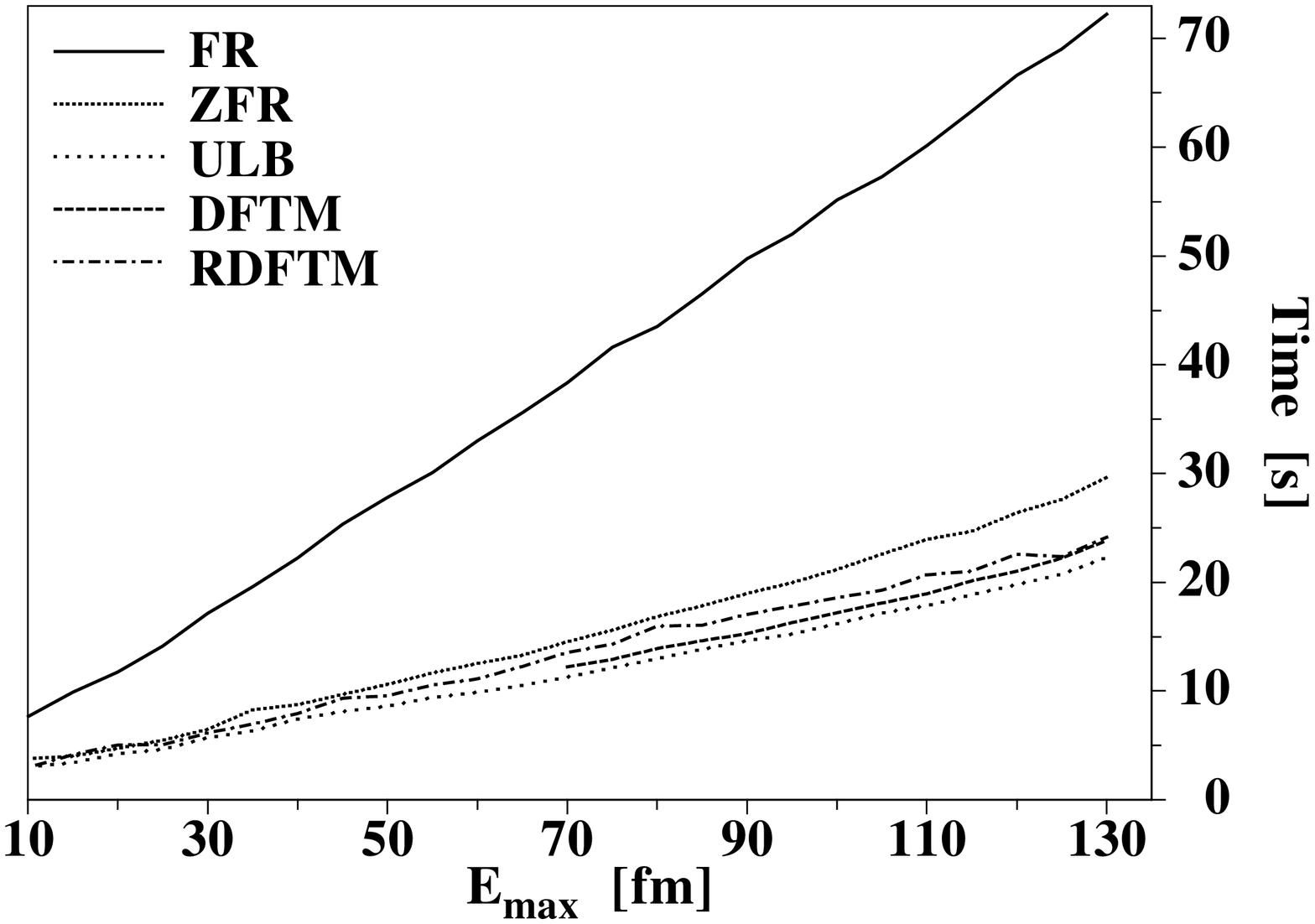}
  \caption{Left panel:  $< \Delta^{n} (N) >_{\kappa}$
  along the Sn isotopic chain. Full dots are experimental five-point odd-even mass
  differences~\cite{audi03a}. Right Panel: CPU time necessary to reach
  convergence in $^{120}Sn$ as a function of $E_{max}$.}
\label{gaps}
\end{figure}

The binding energy $E_{tot}$ is displayed in Fig.~\ref{convener}
for $^{120}Sn$ and $^{170}Sn$ as a function of the energy
$E_{max}$ of the highest HF single-particle state included in the
calculation. For both nuclei, $E_{tot}$ converges significantly
faster for the finite range force FR than for DDDI, regularized
either microscopically (RDFTM and ZFR) or phenomenologically
(DFT)~\footnote{The staggering behavior of $E_{tot}$ as a function
of $E_{max}$ for RDFTM seems to be due to a combined effect of the
sharp cut-off $E_{c}$ used in the corresponding regularization
method and of the discretization of the continuum we perform when
solving the HFB equations. By treating the continuum exactly as
was done in Refs.~\cite{bulgac1,bulgac2}, such a staggering should
disappear.}. If we look at an accuracy of $10$~keV, $E_{tot}
(^{120}Sn)$ is converged around $E_{max} = + 10$~MeV for FR
whereas one has to go up to around $+80 \, MeV$ for ZFR and RDFTM.
The explicit (smooth) cut-off at $E_{c} = 60$~MeV for DFTM makes
necessary to go even slightly higher whereas ULB converges to much
better accuracy for $E_{max} \approx + 5$~MeV thanks to its very
small active window around the Fermi level~\footnote{What is a
converged value depends on the observable and the situation. For
instance, separation or excitation energies converge faster than
absolute binding energies as a function of the basis/box/mesh
sizes or the type of derivative used. On the other hand,
quantities which are not minimized such as pairing gaps or radii
are, at least, as sensitive as binding energies and maybe more in
drip-line nuclei. What is a reasonable converged value also
depends on other sources of numerical inaccuracy. For instance, 3D
codes on the mesh can usually handle a mesh size of 0.8 fm
compared to 0.1/0.25 fm in spherical codes. This can lead to
inaccuracy of the order of $500$~keV in heavy nuclei, even with
the best formula for derivatives~\cite{heenen92a}.}. It is
important to notice, however, that both microscopic regularization
schemes do not motivate the use of such a small active window. Of
course, one has to assess whether this impacts physical
observable. The right panel of Fig.~\ref{gaps} shows that the
intrinsic increase of CPU time needed to tackle the finite-range
force is compensated by its faster convergence as a function of
the basis size. This is a very critical result since the size of
the single-particle basis is (among the size of the box and of the
mesh) what makes calculations on the 3D mesh expensive. With a
mesh of 0.8 fm, 3D codes of the type of
\textsc{ev8}~\cite{bonche05a} can handle 800 wave functions for
systematic calculations performed on supercomputers. For a large
enough box~\cite{duguet06a}, this allows one to go up to $E_{max}
\approx 10/15$~MeV in medium mass nuclei. Depending on the
convergence required, this may rule out some of the pairing
schemes discussed presently. In any case, ULB, FR and ZFR are the
most preferable ones in that respect.

\begin{figure}
  \includegraphics[angle=270,scale=0.31]{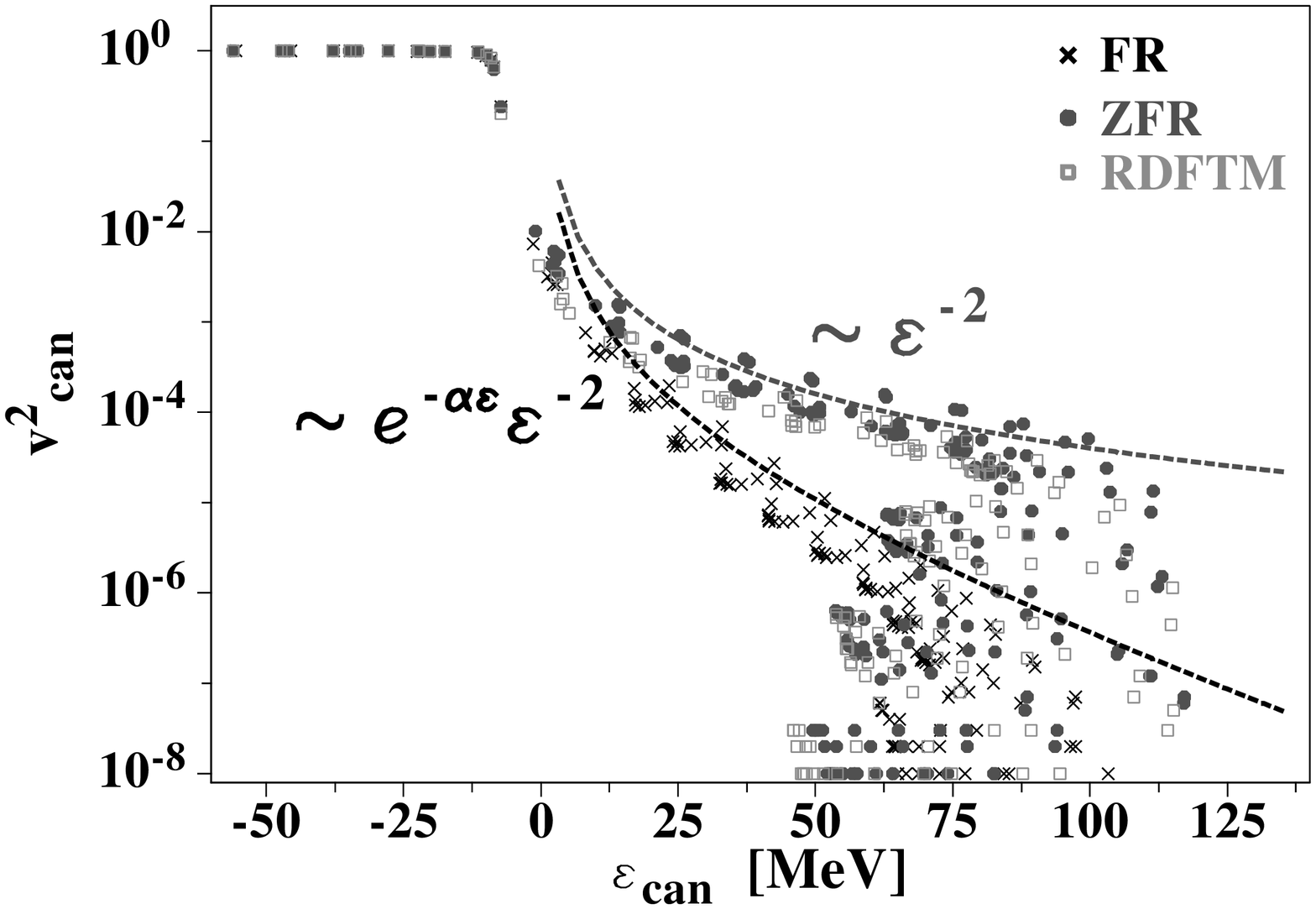} \hspace{-2cm}
  \includegraphics[angle=270,scale=0.31]{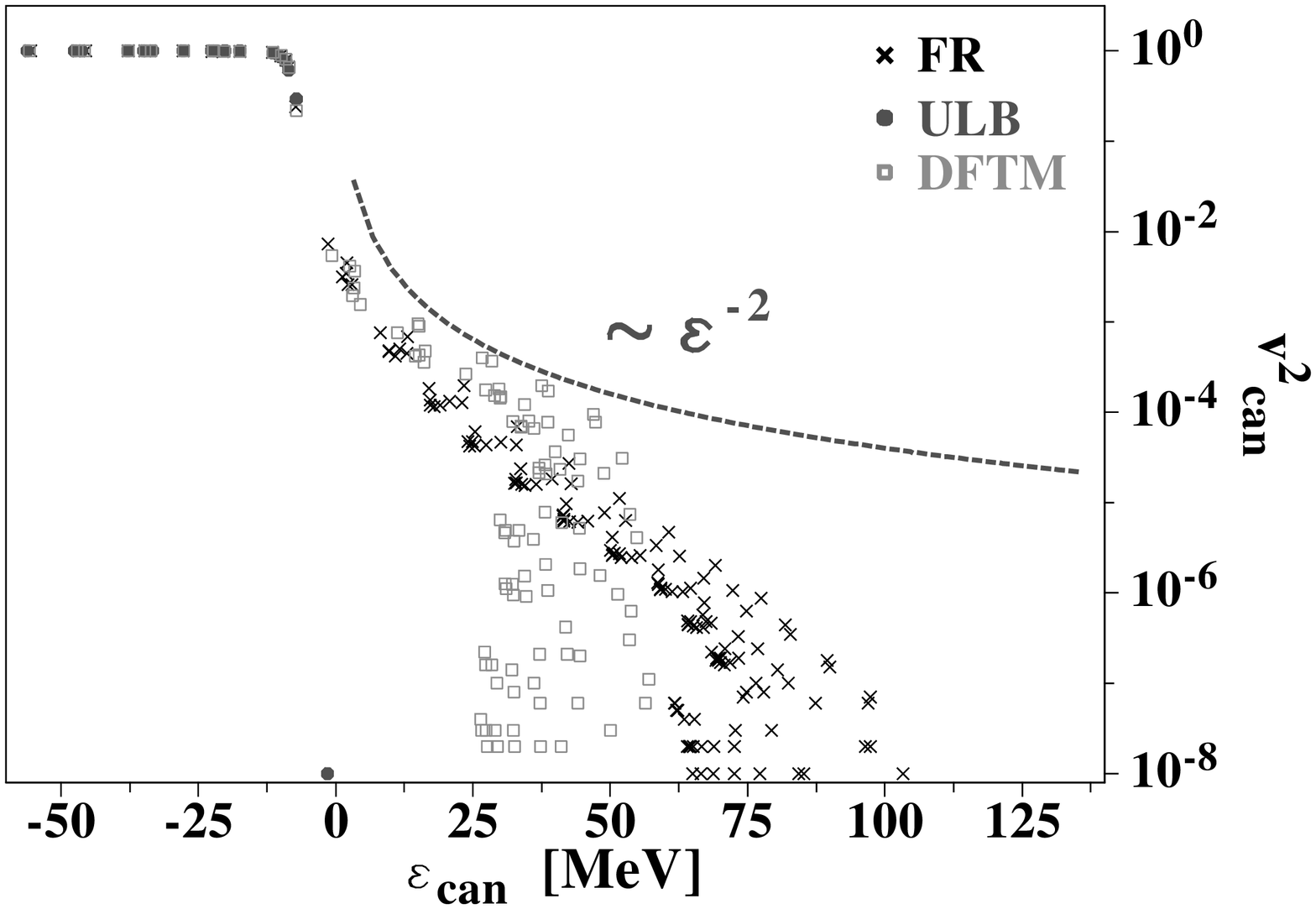}
  \caption{Left panel: $v^{2}_{m}$ for
  $^{120}Sn$. Lines, see text. Right Panel: same with RDFTM
  replacing DFTM and ULB replacing ZFR.}
\label{occ}
\end{figure}

One can nicely relate the convergence properties previously
discussed to occupation of single-particle states in the
continuum. As can be seen from Fig.~\ref{occ}, occupation numbers
decrease much faster for the finite range force FR than for any
DDDI, except for ULB. As a result, physical observable converge
faster for FR than for ZFR, DFTM and RDFTM as a function of the
basis size. Independently on the pairing scheme used, one can
roughly relate the precision required on $E_{tot}$ to the
occupation of single-particle states to be included in the
calculation. For a precision of the order of $10$~keV, one can see
from Fig.~\ref{occ} that states with $v^{2}_{m} > 10^{-4}$ have
been taken into account.

In fact, occupation probabilities reflect directly the properties
of the underlying regularization scheme. Indeed, using a finite
range force or a microscopically regularized DDDI, $v^{2}_{m}$
follows closely the law one can derive for such forces in the
limit $\epsilon_{m} \approx \hbar^2 k^2_{m} / 2m \gg \lambda$.
Such a limit is relevant in finite nuclei because plane-wave
continuum states do not feel the finite nuclear potential at these
energies. The laws obtained in this limit for the bare force used
here~\cite{duguet04a} (thick dashed line in Fig.~\ref{occ}) and
for the corresponding zero-range approximation (thin dashed line
in Fig.~\ref{occ}) read respectively as~\footnote{Single-particle
states which accumulates close to zero on Fig.~\ref{occ}
correspond to canonical states with very small occupation which
are not well converged numerically. Also, $v^{2}_{m}$ patterns
reflect the fact that the calculation is performed for $E_{max} =
+ 130$~MeV.}:

\begin{equation}
v^{2}_{m} \approx \frac{\Delta^{2}}{4 \epsilon^{2}_{m}} \,
e^{-\frac{4 \, m \, \alpha^{2}}{ \hbar^{2}} \, (\epsilon_{m}
-\lambda)} \hspace{1cm} {\rm and} \hspace{1cm} v^{2}_{m} \approx
\frac{\Delta^{2}}{4 \epsilon^{2}_{m}} \, \, \, \, \, .
\end{equation}

On the other hand, occupation numbers associated to DFTM and ULB
do not follow the expected pattern of true zero-range forces and
strongly reflect the properties of the underlying phenomenological
regularization schemes (see the very abrupt decrease close to zero
energy for ULB). Again, the relevant question is whether or not
this influences the description of (pairing) properties of nuclei.

To obtain a first insight, let us focus on DFTM and RDFTM which
only differ through the regularization recipe used. One can see
from Fig.~\ref{convener} that the converged binding energy
provided by the two methods differs by about $300$~keV in
$^{120}Sn$, which is on the edge of being significant as far as
microscopic mass tables are concerned. However, this is quite a
constant shift over a large set of spherical nuclei (see for
example the right panel of Fig.~\ref{convener} for $^{170}Sn$) and
can certainly be easily reabsorbed. As a matter of fact, DFTM and
RDFTM provide identical $S_{2N}$ for all spherical
nuclei~\cite{duguet06a}. Regarding pairing properties,
Fig.~\ref{gaps} shows that average neutron gaps $< \Delta^{n} (N)
>_{\kappa}$ obtained in tin isotopes with the two methods are not
only equal where the forces are fitted but extrapolate in a very
similar way up to the drip-lines. As a conclusion, the
regularization recipe does not seem to matter too much. In order
to tackle the same question regarding a ULB-type regularization, a
parameterization with the same density form-factor as DFTM or
RDFTM should be studied~\cite{duguet06a}.

In any case, one has to be careful before concluding on the
unimportance of employing a microscopic regularization scheme.
Indeed, the conclusion could be different if using a more
realistic density dependence as the one displayed by ZFR, that is,
which reflects the large scattering length of the $NN$ interaction
at low density. Indeed, a DFT-type force was shown to lead to
unrealistic densities and to an unrealistic reduction of the
two-neutron separation energy across the magic number $N=82$ when
used together with such a strongly attractive form factor at low
density. On the other hand, no such unrealistic behavior is seen
with ZFR~\cite{duguet06a} for which the microscopic regularization
method and the strongly attractive form factor at low density have
been derived consistently~\cite{duguet04a}.

There are two elements to the role played by the density
dependence: its spatial character and its isovector nature. Gaps
calculated with ZFR and DFTM are very similar, both in absolute
value and in terms of their isotopic trend~\footnote{The slight
asymmetric behavior between magic numbers of the gaps obtained
with FR and ZFR is due to the "asymmetric option" of the recast
used and to the approximation associated to it when going from
infinite matter to finite nuclei. This can be understood by
employing the exact bare force directly or by performing the
recast of the gap equation directly in the finite
nucleus~\cite{lesinski06a}.}. This is due to the behavior of ZFR
between volume and surface which confirmed the phenomenological
findings of Refs.~\cite{duguet2,doba4,doba2,doba7}. Surface-peaked
DDDI on the other hand provide too strong pairing, especially in
neutron rich nuclei. This is clear when comparing DFTM and DFTS
which differ only through the spatial character of $f
\left(\vec{r}\right)$. The increasing difference between DFTM and
DFTS as we move towards the drip-line reflects in fact mainly the
isospin character associated with their dependence on $\rho_{0}$.
Forces depending on $\rho_{0}$ only give too strong pairing
correlations in neutron rich matter as explained in
Ref.~\cite{duguet05b}~\footnote{Such a statement deals with
pairing correlations generated by the bare $NN$ force. On the
other hand, the isospin character of beyond mean-field effects is
far from being understood.}. The latter effect being proportional
to how much the density form-factor varies over the volume of the
nucleus, it is much more pronounced for DFTS than for DFTM. The
rather weak dependence of DFTM on $\rho_{0}$ explains why the
pairing gaps predicted by DFTM and ZFR towards the drip-line
remains similar and are not spoiled by their different isospin
nature.

\section{Conclusions}
\label{conclusions}

In the present paper, we briefly compare the main pairing schemes
currently used in HFB calculations of nuclei (except for the Gogny
force). We employ spherical 1D HFB calculations of tin isotopes to
assess the impact of the regularization recipes and of the density
dependence used in connection with zero-range forces. Beyond usual
phenomenological DDDI, the recently proposed microscopic pairing
force equivalent to the bare $NN$ force is
discussed~\cite{duguet04a}. It was shown that only the latter
finite-range force and phenomenological DDDI using small active
pairing windows around the Fermi energy can currently be tackled
in calculations on a 3D mesh.

The results presented here constitute a preview of a more
extensive and systematic study of existing phenomenological and
microscopic pairing schemes~\cite{duguet06a}. In addition to
analyzing the formal and technical differences between those
schemes, the forthcoming study will focus on the impact these
differences may or may not have on physical properties of nuclear
ground-states.

\section{Acknowledgments}
\label{secremer}

The authors thank M. Bender for fruitful discussions and for a
careful reading of the manuscript.

\bibliographystyle{aipproc}

\bibliography{kyoto05}

\IfFileExists{\jobname.bbl}{}
 {\typeout{}
  \typeout{******************************************}
  \typeout{** Please run "bibtex \jobname" to optain}
  \typeout{** the bibliography and then re-run LaTeX}
  \typeout{** twice to fix the references!}
  \typeout{******************************************}
  \typeout{}
 }

\end{document}